\documentclass{article}
\usepackage{ijcai18}
\usepackage{booktabs} 
\usepackage{times}
\usepackage{stfloats}
\usepackage{xcolor}
\usepackage{soul}
\usepackage{bbding}
\usepackage{pifont}
\usepackage[utf8]{inputenc}
\usepackage{mathtools}
\usepackage{comment}
\usepackage[small]{caption}

\title{NeuRec: On Nonlinear Transformation for Personalized Ranking}

\author{
Shuai Zhang$^1$,
Lina Yao$^1$,
Aixin Sun$^2$,
Sen Wang$^3$,\\
\textbf{Guodong Long}$^4$\textbf{,}
and \textbf{Manqing Dong}$^1$
\\
$^1$ University of New South Wales,\\
$^2$ Nanyang Technological University, \\$^3$ Griffith University\\
$^4$ University of Technology Sydney\\
\{shuai.zhang@student., lina.yao@\}unsw.edu.au
}
\begin{document}

\maketitle

\begin{abstract}
Modeling user-item interaction patterns is an important task for personalized recommendations. Many recommender systems are based on the assumption that there exists a linear relationship between users and items while neglecting the intricacy and non-linearity of real-life historical interactions. In this paper, we propose a neural network based recommendation model (NeuRec) that untangles the complexity of user-item interactions and establish an integrated network to combine non-linear transformation with latent factors. We further design two variants of NeuRec: user-based NeuRec and item-based NeuRec, by focusing on different aspects of the interaction matrix. Extensive experiments on four real-world datasets demonstrated their superior performances on personalized ranking task.


\end{abstract}

\section{Introduction}

Recommender systems have been playing  a critical role in the realms of retail, social networking, and entertainment industries. Providing personalized recommendations is an important commercial strategy for online websites and mobile applications. There are two major recommendation tasks: rating prediction and personalized ranking. The former usually needs explicit ratings(e.g., 1-5 stars) while the latter aims to generate a ranked list of items in descending order based on the estimated preferences for each user. In many real world scenarios where only implicit feedback is available, personalized ranking is a more appropriate and popular choice ~\cite{rendle2009bpr}. Collaborative filtering (CF) is a \textit{de facto} approach which has been widely used in many real-world recommender systems~\cite{ricci2015recommender}. CF assumes that user-item interactions can be modelled by inner product of user and item latent factors in a low-dimensional space.  An effective and widely adopted ranking model based on CF is Bayesian Personalized Ranking (BPR)~\cite{rendle2009bpr} which optimizes the ranking lists with a personalized pairwise loss. Another state-of-the-art model is sparse linear method (SLIM)~\cite{6137254} which recommends top-$n$ items via sparse linear regression. While BPR and SLIM have been shown to perform well on ranking task, we argue that they are hindered by a critical limitation: both of them are built on the assumption that there exists a linear relationship between users and items, while the relationship shall be more complex in real-life scenarios.


In recent years, researchers have demonstrated the efficacy of deep neural model for recommendation problems~\
\cite{zhang2017deep,Karatzoglou:2017:DLR:3109859.3109933}. Deep neural network can be integrated into classic recommendation models such as collaborative filtering~\cite{he2017neural,Tay:2018:LRM:3178876.3186154} and content based approaches~\cite{cheng2016wide,DBLP:journals/corr/abs-1801-09251} to enhance their performances.  Many deep neural techniques such as multi-layered perceptron (MLP), autoencoder (AE), recurrent neural network (RNN) and convolutional neural network (CNN) can be applied to recommendation models.  AE is usually used to incorporate side information of users/items. For example, ~\cite{wang2015collaborative} and ~\cite{Zhang:2017:AEH:3077136.3080689}  proposed integrated models by combining latent factor model (LFM) with different variants of autoencoder; AE can also be adopted to reconstruct the rating matrix directly ~\cite{sedhain2015autorec}. CNN is mainly used to extract features from textual~\cite{kim2016convolutional,zheng2017joint},  audio~\cite{van2013deep} or visual~\cite{he2016vbpr} content. RNN can be used to model the sequential patterns of rating data or session-based recommendation~\cite{hidasi2015session}. For example, ~\cite{Wu:2017:RRN:3018661.3018689} designed a recurrent neural network based rating prediction model to capture the temporal dynamics of rating data; ~\cite{hidasi2015session} proposed using RNN to capture the interconnections between sessions. Some works attempted to generalize traditional recommendation models into neural versions. For example, ~\cite{he2017neural,He:2017:NFM:3077136.3080777} designed the neural translations of LFM and factorization machine to model user-item interactions; ~\cite{ijcai2017-447} proposed a deep matrix factorization model to anticipate user's preferences from historical explicit feedback.

Most previous works focused upon either explicit feedback (rating prediction task) or representation learning from abundant auxiliary information instead of interpreting user-item relationships in depth. In this work, we aim to model the user-item intricate relationships from implicit feedback, instead of explicit ratings, by applying multi-layered nonlinear transformations. The main contributions are as follows:

\begin{itemize}
    \item We propose two recommendation models with deep neural networks, user-based NeuRec (U-NeuRec) and item-based NeuRec (I-NeuRec), for personalized ranking task. We present an elegant integration of LFM and neural networks which can capture both the linearity and non-linearity in real-life datasets.
    \item With deep neural networks, we managed to reduce the number of parameters of existing advanced models while achieving superior performances.
\end{itemize}
\section{Preliminaries}
To make this paper self-contained, we first define the research problem and introduce two highly relevant previous works.
\subsection{Problem Statement}
Let $M$ and $N$ denote the total number of users and items in a recommender system, so we have a $M \times N$ interaction matrix $X \in \mathcal{R}^{M \times N}$. We use low-case letter $u  \in \{1,...,M\}$ and $i \in \{1,...,N\}$ to denote user $u$ and item $i$ respectively, and $X_{ui}$ represents the preference of user $u$ to item $i$. In our work, we will use two important vectors: $X_{u*}$ and $X_{*i}$. $X_{u*} = \{X_{u1},...,X_{uN}\}$ denotes user $u$'s preferences toward all items; $X_{*i} = \{X_{1i},...,X_{Mi}\}$ means the preferences for item $i$ received from all users in the system. We will focus on recommendation with implicit feedback here. Implicit feedback such as, click, browse and purchase is widely accessible and easy to collect. We set $X_{ui}$ to $1$ if the interaction between user $u$ and item $i$ exists, otherwise, $X_{ui}=0$. Here, $0$ does not necessarily mean user $u$ dislikes item $i$, it may also mean that the user does not realize the existence of item $i$.


\subsection{Latent Factor Model}
Latent factor model (LFM) is an effective methodology for model-based collaborative filtering. It assumes that the user-item affinity can be derived from low-dimensional representations of users and items. Latent factor method has been widely studied and many variants have been developed~\cite{Koren:2009:MFT:1608565.1608614,koren2008factorization,Zhang:2017:AEH:3077136.3080689,Salakhutdinov:2007:PMF:2981562.2981720}. One of the most successful realizations of LFM is matrix factorization. It factorizes the interaction matrix into two low-rank matrices with the same latent space of dimensionality $k$ ($k$ is much smaller than $M$ and $N$), such that user-item interactions are approximated as inner product in that space
\begin{equation}
    X_{ui} = U_u \cdot V_i
\end{equation}
where $U \in \mathcal{R}^{M \times k}$ is the user latent factor and $V \in \mathcal{R}^{N \times k}$ is the item latent factor. With this low rank approximation, it compresses the original matrix down to two smaller matrices.

\subsection{Sparse Linear Method}

SLIM~\cite{6137254} is a sparse linear model for top-$n$ recommendation. It aims to learn a sparse aggregation coefficient matrix $S \in \mathcal{R}^{N \times N}$. $S$ is reminiscent of the similarity matrix in item-based neighbourhood CF (itemCF)~\cite{1167344}, but SLIM learns the similarity matrix as a least squares problem rather than determines it with predefined similarity metrics (e.g., cosine, Jaccard etc.). It finds the optimal coefficient matrix $S$ by solving the following optimization problem
\begin{equation*}
  \begin{aligned}
    \underset{S}{min}\parallel X - XS\parallel_F^2 + \lambda \parallel S \parallel_F^2 + \mu \parallel S\parallel_1 \\
s.t.  S\geq 0, diag(S) = 0
\end{aligned}
\label{slim}
\end{equation*}
The constraints are intended to avoid trivial solutions and ensure positive similarities. The $\ell_1$ norm is adopted to introduce sparsity to matrix $S$. SLIM can be considered as a special case of LFM with $X \Leftrightarrow U$ and $S \Leftrightarrow V$. SLIM is demonstrated to outperform numerous models in terms of top-$n$ recommendation. Nevertheless, we argue that it has two main drawbacks: (1) From the definition, the size of $S$ is far larger than the two latent factor models, that is,  $N \times N \gg (N \times k + M \times k )$, which also results in higher model complexity. Even though it can be improved via feature selection by first learning an itemCF model, this sacrifices model generalization as it heavily relies on other pre-trained recommendation models; (2) SLIM assumes that there exists strong linear relationship between interaction matrix and $S$. However, this assumption does not necessarily holds. Intuitively, the relationship shall be far more complex in real world applications due to the dynamicity of user preferences and item changes. In this work, we aim to address these two problems. Inspired by LFM and recent advances of deep neural network on recommendation tasks, we propose employing a deep neural network to tackle the above disadvantages by introducing non-linearity to top-$n$ recommendations.

\section{Proposed Methodology}
In this section, we present a novel nonlinear model based on neural network for top-$n$ recommendation and denote it by \textbf{NeuRec}. Unlike SLIM which directly applies linear mapping on the interaction matrix $X$, NeuRec first maps $X$ into a low-dimensional space with multi-layer neural networks. This transformation not only reduces the parameter size, but also incorporates non-linearity to the recommendation model. Then the user-item interaction is modeled by inner product in the low-dimensional space. Based on this approach, we further devise two variants, namely, U-NeuRec and I-NeuRec.

\subsection{User-based NeuRec}
\begin{figure*}
\centering
\includegraphics[width=0.85\textwidth]{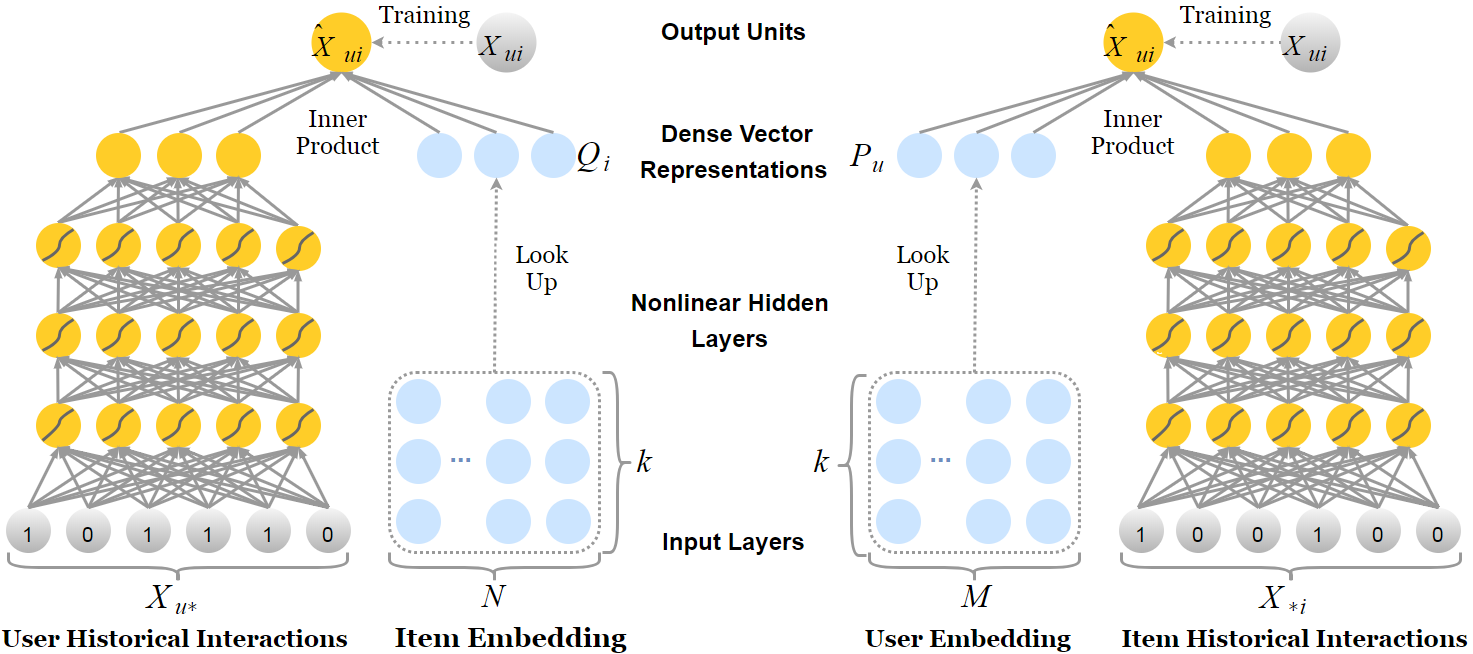}
\caption{Illustration of User-based NeuRec (left) and item-based NeuRec(right). Both of them has two parts: a multi-layer perceptron with $X_{u*}$ (or $X_{*i}$) as input and item (or user) latent factor.}
\label{nlrec}
\vspace{-4mm}
\end{figure*}

For user-based NeuRec, we first get the high-level dense representations from the rows of $X$ with feed-forward neural networks. Note that $X$ is constructed with training data, so there are no leakages of test data in this model. Let $W_j$ and $b_j$, $j=\{1,...,L\}$ ($L$ is the number of layers) denote the weights and biases of layer $j$. For each user, we have
\begin{equation*}
  \begin{aligned}
h_1(X_{u*}) &= f(W_1 X_{u*} + b_1) \\
h_j(X_{u*}) &= f(W_j h_{j-1} + b_j) \\
h_L(X_{u*}) &= f(W_L h_{L-1} + b_L)
\end{aligned}
\end{equation*}
where $f(\cdot)$ is a non-linear activation function such as $sigmoid$, $tanh$ or $relu$.  The dimension of output $h_L(X_{u*})$ is usually much smaller than original input $X_{u*}$. Suppose the output dimension is $k$ (we reuse the latent factor size $k$ here), we have an output $h_L(X_{u*}) \in \mathcal{R}^{k}$ for each user. Same as latent factor models, we define an item latent factor $Q_i \in \mathcal{R}^{k}$ for each item, and consider $h_L(X_{u*})$ as user latent factor.  The recommendation score is computed by the inner product of these two latent factors
\begin{equation}
    \hat{X_{ui}} = h_L(X_{u*}) \cdot Q_i
    \label{unfm}
\end{equation}
To train this model, we minimize the regularized squared error in the following form
\begin{equation}
    \underset{W*,Q*, b*}{min}\sum_{ u, i }(X_{ui} -\hat{X_{ui}})^2 + \lambda ( \parallel W \parallel_F^2 + \parallel Q \parallel_F^2)
    \label{opt}
\end{equation}
Here,  $\lambda$ is the regularization rate. We adopt the Frobenius norm to regularize weight $W$ and item latent factor $Q$. Since parameter $Q$ is no longer a similarity matrix but latent factors in a low-dimensional space, the constraints in SLIM and $\ell_1$ norm can be relaxed.
For optimization, we apply the Adam algorithm~\cite{kingma2014adam} to solve  this objective function. Figure \ref{nlrec}(left) illustrates the architecture of U-NeuRec.

\subsection{Item-based NeuRec}
Likewise, we use the column of $X$ as input and learn a dense representation for each item with a multi-layered neural network
\begin{align}
h_1(X_{*i}) &= f(W_1 X_{*i} + b_1) \\
h_j(X_{*i}) &= f(W_j h_{j-1} + b_j) \\
h_L(X_{*i}) &= f(W_L h_{L-1} + b_L)
\end{align}
Let $P_u$ denote the user latent factor for user $u$, then the preference score of user $u$ to item $i$ is computed by
\begin{equation}
    \hat{X_{ui}} = P_u \cdot h_L(X_{*i})
    \label{infm}
\end{equation}
We also employ a regularized squared error as the training loss. Thus, the objective function of item-based NeuRec is formulated as
\begin{equation}
    \underset{W*,P*, b*}{min}\sum_{ u, i }(X_{ui} -\hat{X_{ui}})^2 + \lambda ( \parallel W \parallel_F^2 + \parallel P \parallel_F^2)
\end{equation}
the optimal parameters can also be learned with Adam Optimizer as well. The architecture of I-NeuRec is illustrated in Figure \ref{nlrec}(right).

\subsection{Dropout Regularization}
Dropout~\cite{srivastava2014dropout} is an effective regularization technique for neural networks. It can reduce the co-adaptation between neurons by randomly dropping some neurons during training. Unlike traditional dropout which is usually applied on hidden layers, here, we propose applying the dropout operation on the input layer $X_{u*}$ or $X_{i*}$ (We found that the improvement of applying the dropout on hidden layers is subtle in our case). By randomly dropping some historical interactions, we could prevent the model from learning the identity function and increase the robustness of NeuRec.

\subsection{Relation to LFM and SLIM}
In this section, we shed some light on the relationships between NeuRec and LFM / SLIM. NeuRec can be regarded as a neural integration of LFM and sparse linear model. NeuRec utilizes the concepts of latent factor in LFM. The major difference is that either item or user latent factors of NeuRec are learned from the rating matrix with deep neural network. In addition, NeuRec also manages to capture both negative and positive feedback in an integrated manner with rows or columns of $X$ as inputs. To be more precise, U-NeuRec is a neural extension of SLIM. If we set $f$ to identity function and enforce $W$ to be a uniform vector of 1 and omit the biases, we have $h_L(X_{u*})^T \Leftrightarrow X_{u*}$. Hence, U-NeuRec will degrade to a SLIM with $S \Leftrightarrow Q$. Note that the sparsity and non-negativity constraints are dropped. I-NeuRec has no direct relationship with SLIM. Nonetheless, it can be viewed as a symmetry version of U-NeuRec. Since the objective functions of NeuRec and SLIM are similar, the complexities of these two models are linear to the size of the interaction matrix. Yet, NeuRec has less model parameters.

\subsection{Pairwise Learning Approach}
NeuRec can be boiled down to a pairwise training scheme with Bayesian log loss.
\begin{equation}
\begin{multlined}
    \underset{\Theta}{min} \sum_{(u,i^+, i^-)} - log(\sigma(\hat{X}_{ui^+} - \hat{X}_{ui^-}))  + \Omega(\Theta)
    \label{pairwise}
\end{multlined}
\end{equation}
Where $\Theta$ is the model parameters, $\Theta = \{W*,Q*, b*\}$ for U-NeuRec, and $\Theta = \{W*, P*, b*\}$ for I-NeuRec; $\Omega$ is Frobenius regularization; $i^+$ and $i^-$ represent observed and unobserved items respectively. The above pairwise method is intended to maximize the difference between positive items and negative items. However, previous studies have shown that optimizing these pairwise loss does not necessarily lead to best ranking performance~\cite{Zhang:2013:OTC:2484028.2484126}. To overcome this issue, we adopt a non-uniform sampling strategy: in each epoch, we randomly sampled $t$ items from negative samples for each user, calculate their ranking score and then treat the item with the highest rank as the negative sample. The intuition behind this algorithm is that we shall rank all positives samples higher than negatives samples.

\section{Experiments}
In this section, we conduct experiments on four real-world datasets and analyze the impact of hyper-parameters.


\subsection{Experimental Setup}

\subsubsection{Datasets Description}
We conduct experiments on four real-world datasets: Movielens HetRec, Movielens 1M, FilmTrust and Frappe.  The two Movielens datasets\footnote{https://grouplens.org/datasets/movielens/} are collected by GroupLens research\cite{Harper:2015:MDH:2866565.2827872}. Movielens HetRec is released in HetRec 2011\footnote{http://recsys.acm.org/2011 }. It consists of $855598$ interactions from $10109$ movies and $2113$ users. They are widely used as benchmark datasets for evaluating the performance of recommender algorithms.  FilmTrust is crawled from a movie sharing and rating website by Guo et al.~\cite{guo2013novel}. Frappe~\cite{baltrunas2015frappe} is an Android application recommendation dataset which contains around a hundred thousand records from $957$ users on over four thousand mobile applications. The interactions of these four datasets are binarized  with the approach introduced in Section 2.1.

\subsubsection{Evaluation Metrics}
To appropriately evaluate the overall performance for ranking task, the evaluation metrics include Precision and Recall with different cut-off value (e.g., P@5, P@10, R@5 and R@10), Mean Average Precision (MAP), Mean Reciprocal Rank (MRR) and Normalized Discounted Cumulative Gain (DNCG). These metrics are used to evaluate the quality of recommendation lists regarding different aspects~\cite{liu2009learning,shani2011evaluating}: Precision, Recall and MAP are used to evaluate the recommendation accuracy, as they only consider the hit numbers and ignore the rank positions; MRR and DNCG are two rank-aware measures with which higher ranked positive items are prioritized, thus they are more suitable for assessing the quality of ranked lists. We omit the details for brevity.

\subsection{Implementation Details}

We implemented our proposed model based on Tensorflow\footnote{https://www.tensorflow.org/} and tested it on a NVIDIA TITAN X Pascal GPU. All models are learned with min-batch Adam. We do grid search to determine the hyper-parameters. For all the datasets, we implement a five hidden layers neural network  with constant structure for the neural network part of NeuRec and use \textit{sigmoid} as the activation function. For ML-HetRec, we set the neuron number of each layer to $300$, latent factor dimension $k$ to $50$ and dropout rate to $0.03$; For ML-1M, neuron number is set to $300$, $k$ is set to $50$, and dropout rate is set to $0.03$. The neuron size for FilmTrust is set to $150$ and $k$ is set to $40$. We do not use dropout for this dataset; For Frappe, neuron size is set to $300$, $k$ is set to $50$ and dropout rate is set to $0.03$.  We set the learning rate to $1e-4$ for ML-HetRec, ML-1M and Frappe. The learning rate for FilmTrust is $5e-5$. For ML-HetRec, ML-1M and FilmTrust, we set the regularization rate to $0.1$, and that for Frappe is set to $0.01$. For simplicity, we adopt the same parameter setting for pairwise training method. We use 80\% user-item pairs as training data and hold out 20\% as the test set, and estimate the performance based on five random train-test splits.

\subsection{Results and Discussions}

Since NeuRec is designed to overcome the drawbacks of LFM and \textbf{SLIM}, so they are two strong baselines for comparison to demonstrate if our methods can overcome their disadvantages. Specifically, we choose \textbf{BPRMF}~\cite{rendle2009bpr}, a personalized ranking algorithm based on matrix factorization, as the representative of latent factor model. Similar to~\cite{6137254},  we adopt neighbourhood approach to accelerate the training process of SLIM.
For fair comparison, we also report the results of \textbf{mostPOP} and two neural network based models: \textbf{GMF} and \textbf{NeuMF}~\cite{he2017neural}, and follow the configuration proposed in ~\cite{he2017neural}. The recent work DMF~\cite{ijcai2017-447} is tailored for explicit datasets and not suitable for recommendations on implicit feedback, so it is unfair to compare our method with it.

\subsubsection{Parameter Size}

The parameter size of SLIM is $N \times N$, while I-NeuRec has $S_{nn} + k \times M$ parameters and U-NeuRec has $S_{nn} + k \times N$. $S_{nn}$ is the size of the neural network. Usually, our model can reduce the number of parameters largely (up to 10 times).


\subsubsection{Overall Comparisons}
\begin{table*}[t!]
\centering

\label{result}
\begin{tabular}{l|cccccc}
\toprule
\midrule
\multicolumn{7}{c}{\textbf{MOVIELENS HetRec}} \\
\midrule
Methods& Precision@5 & Precision@10 & Recall@5 & Recall@10 & MAP & MRR \\
\midrule
mostPOP  & $0.455 \pm 0.002$ & $0.403 \pm 0.003$  & $0.042 \pm 0.001$  &  $ 0.070   \pm 0.001$  & $0.181 \pm 0.001$ & $0.651 \pm 0.004$\\

BPRMF      & $0.537 \pm 0.002$ & $0.486 \pm 0.001$&
$0.052 \pm 0.001$   &  $0.090 \pm 0.001 $ &$0.246 \pm 0.001$ & $0.713 \pm 0.001$ \\
GMF    & $0.540\pm0.002  $& $ 0.487\pm 0.001 $    &$ 0.053 \pm 0.001$    &$ 0.090\pm0.001 $  & $0.248 \pm 0.001$& $ 0.719\pm 0.005$ \\

SLIM    & $0.528 \pm 0.002 $& $0.465 \pm 0.002$    &$0.055  \pm 0.001$    &$0.090 \pm 0.001$  & $ 0.227 \pm 0.001$& $0.755 \pm 0.001$ \\
NeuMF    & $0.535\pm0.006  $& $ 0.485\pm 0.004 $    &$ 0.053 \pm 0.001$    &$ 0.091\pm0.001 $  & $0.248 \pm 0.002$& $ 0.722\pm 0.006$ \\
\midrule
I-NeuRec     &  $\textbf{0.603} \pm  \textbf{0.004}$ & $\textbf{0.542} \pm \textbf{0.003}$  & $\textbf{0.060} \pm \textbf{0.001}$ & $\textbf{0.101} \pm \textbf{0.001}$  & $\textbf{0.278} \pm \textbf{0.002}$   &$\textbf{0.772} \pm \textbf{0.006}$ \\

U-NeuRec     &  $\underline{0.601}\pm 0.004  $& $\underline{0.538}\pm 0.004  $& $\underline{0.059}\pm 0.001 $ & $\underline{0.098}\pm 0.002  $ &$\underline{0.271}\pm 0.002$ &    $\underline{0.768}\pm 0.003$    \\

\midrule
\multicolumn{7}{c}{\textbf{MOVIELENS 1M}} \\
\midrule
Methods& Precision@5 & Precision@10 & Recall@5 & Recall@10 & MAP & MRR \\
\midrule
mostPOP& $0.210 \pm 0.001$& $0.182 \pm 0.002$ & $0.041 \pm 0.001$   &  $0.066  \pm 0.001 $    & $0.102 \pm 0.001 $& $0.392 \pm 0.004 $\\
BPRMF    &  $0.354 \pm 0.003 $& $0.307 \pm 0.001$& $0.078  \pm 0.001 $ & $0.130 \pm 0.001$ & $0.199 \pm 0.001$ & $0.572 \pm 0.003 $   \\
GMF   & $0.367\pm0.001  $& $ 0.316\pm 0.001 $    &$ 0.081 \pm 0.001$    &$ 0.134\pm0.001 $  & $0.201 \pm 0.001$& $ 0.589\pm 0.006$ \\
SLIM      &$0.340 \pm 0.004 $ & $0.291 \pm 0.002$& $0.091 \pm 0.001$ & $0.148 \pm 0.001$ & $0.198 \pm 0.001$ & $0.585 \pm 0.003$  \\
NeuMF    & $0.367\pm 0.004 $& $ 0.319\pm 0.002$    &$ 0.081 \pm0.002 $    &$ 0.135\pm 0.002$  & $ 0.208 \pm 0.002 $& $ 0.586\pm0.002 $ \\
\midrule
I-NeuRec     &$\underline{0.414} \pm 0.001$  & $\underline{0.359} \pm 0.001$ &  $\underline{0.100} \pm 0.001$ & $\underline{0.161} \pm 0.001$&  $\underline{0.242} \pm 0.001$&   $\underline{0.636} \pm 0.003$     \\
U-NeuRec    & $\textbf{0.419} \pm \textbf{0.002}$  & $\textbf{0.362} \pm \textbf{0.003}$ & $\textbf{0.103} \pm\textbf{0.001}$ & $\textbf{0.165} \pm \textbf{0.002}$ & $\textbf{0.245} \pm \textbf{0.002}$& $\textbf{0.650} \pm \textbf{0.003}$     \\
\midrule
\multicolumn{7}{c}{\textbf{FILMTRUST}} \\
\midrule
Methods& Precision@5 & Precision@10 & Recall@5 & Recall@10 & MAP & MRR \\
\midrule
mostPOP & $0.418 \pm 0.004$ & $0.350 \pm 0.002 $ & $0.397 \pm 0.008$& $\underline{0.631}  \pm 0.004 $  &  $0.489 \pm 0.002$ & $ 0.618 \pm 0.004$\\
BPRMF     & $0.412 \pm 0.005$& $0.347 \pm 0.000$ & $0.391 \pm 0.009 $& $0.613 \pm 0.007$& $0.476 \pm 0.004$&$0.600 \pm 0.007 $     \\
GMF    & $0.393\pm0.004  $& $ 0.342\pm 0.003 $    &$ 0.393 \pm 0.004$    &$ 0.608\pm0.002 $  & $0.481 \pm 0.004$& $ 0.613\pm 0.008$ \\
SLIM        & $ \underline{0.431} \pm 0.002$ & $\underline{0.352} \pm 0.002$ &  $\underline{0.422} \pm 0.005$ & $0.625 \pm 0.003$ & $\underline{0.507} \pm 0.003$ &  $\underline{0.647} \pm 0.002$ \\
NeuMF    & $0.413\pm 0.003 $& $  0.350\pm0.003$   &$ 0.392\pm 0.002$    &$ 0.626 \pm 0.007$   & $ 0.483\pm0.001$& $ 0.609\pm0.005 $ \\
\midrule
I-NeuRec         &  $ 0.421\pm0.005 $ &  $ 0.347\pm0.002 $ & $0.405 \pm 0.011$ &$ 0.619\pm0.005 $ & $ 0.491\pm0.008 $ &   $ 0.621\pm0.012  $   \\
U-NeuRec       & $\textbf{0.441} \pm \textbf{0.003}$   & $\textbf{0.358} \pm \textbf{0.002}$  & $\textbf{0.446} \pm \textbf{0.004}$& $\textbf{0.654} \pm \textbf{0.007} $ & $\textbf{0.530} \pm \textbf{0.006}$ &  $\textbf{0.667} \pm \textbf{0.008}$ \\

\midrule
\multicolumn{7}{c}{\textbf{FRAPPE}} \\
\midrule
Methods& Precision@5 & Precision@10 & Recall@5 & Recall@10 & MAP & MRR \\
\midrule
mostPOP &$0.034 \pm 0.001$ & $0.026 \pm 0.001$ &$0.054 \pm 0.001$& $0.075 \pm 0.001$ & $0.041 \pm 0.002$& $0.115 \pm 0.001$ \\
BPRMF     & $0.055 \pm 0.003$ & $0.052 \pm 0.003$ & $0.059 \pm 0.002$  & $0.095 \pm 0.005$& $0.052 \pm 0.002$  &  $0.134 \pm 0.005$  \\

GMF    & $0.055 \pm 0.004$ &  $0.043 \pm 0.002 $  & $0.066 \pm 0.005$&  $0.095 \pm 0.006$& $0.094 \pm 0.001$&$0.151 \pm 0.001$ \\
SLIM        & $0.089 \pm 0.003$ & $0.064 \pm 0.001$ & $0.065 \pm 0.003$ & $0.092 \pm 0.003$& $\underline{0.108} \pm 0.003$ & $\underline{0.195} \pm 0.003$ \\
NeuMF    & $0.072\pm 0.002 $& $ 0.056\pm0.002 $    &$  \underline{0.076}\pm 0.002$    &$ \textbf{0.105}\pm\textbf{0.004}$  & $ 0.104\pm 0.002$& $ 0.174\pm 0.004$ \\
\midrule
I-NeuRec        & $\textbf{0.106} \pm \textbf{0.003}$   & $\textbf{0.075} \pm \textbf{0.001}$  & $\textbf{0.078} \pm \textbf{0.003}$& $\underline{0.102} \pm 0.005 $ & $\textbf{0.125} \pm \textbf{0.004}$ &  $\textbf{0.211} \pm \textbf{0.006}$ \\
U-NeuRec     & $\underline{0.093}\pm 0.006 $  & $\underline{0.068}\pm 0.003 $  & $0.067\pm 0.007 $ & $0.094\pm 0.006 $ &   $0.107 \pm 0.004$  & $0.185\pm 0.002$ \\
\midrule
\bottomrule
\end{tabular}
\caption{Precision@5, Precision@10, Recall@5, Recall@10, MAP and MRR comparisons on Movielens HetRec, Movielens 1M, FilmTrust, Frappe. Best performance is in boldface and second best is underlined. I-NeuRec and U-NeuRec are models proposed by us.}
\vspace{-3mm}
\label{resultsss}
\end{table*}

Table \ref{resultsss} and Figure 2 summarize the overall performance of baselines and NeuRec. From the comparison, we can observe that our methods constantly achieve the best performances on these four datasets not only in terms of prediction accuracy but also ranking quality. Higher MRR and NDCG mean that our models can effectively rank the items user preferred in top positions. Performance gains of NeuRec over the best baseline are: Movielens HetRec ($8.61\%$), Movielens 1M ($12.29\%$), FilmTrust ($3.43\%$), Frappe ($8.93\%$).  The results of I-NeuRec and U-NeuRec are very close and better than competing baselines.  The subtle difference between U-NeuRec and I-NeuRec might be due to the distribution differences of user historical interactions and item historical interactions (or the number of users and items). We found that the improvement of NeuMF over GMF are not significant, which might be due to the overfitting caused by the use of dual embedding spaces~\cite{Tay:2018:LRM:3178876.3186154}. Although the improvements of pairwise based U-NeuRec and I-NeuRec are subtle (in Tables 2 and 3), they are still worth being investigated. From the results, we observe that U-NeuRec is more suitable for pairwise training. In U-NeuRec, positive item and negative item are represented by two independent vectors $Q_{i^+}$ and $Q_{i^-}$, while in I-NeuRec, they need to share the same network with input $X_{*i^+}$ or $X_{*i^-}$. Therefore, the negative and positive samples will undesirably influence each other.

\begin{table}[h]
\centering

\label{my-label}
\begin{tabular}{c|cccc}
\toprule
 & \multicolumn{1}{c}{ML HetRec}     & \multicolumn{1}{c}{ML 1M}       & \multicolumn{1}{c}{FilmTrust}   & \multicolumn{1}{c}{FRAPPE}   \\
\midrule
P@5     & 0.521 & 0.347 & 0.418  &  0.038\\
P@10    & 0.473  & 0.303     &  0.349  & 0.032 \\
R@5     & 0.047   & 0.077   & 0.402 & 0.054   \\
R@10    &  0.082 & 0.128   &  0.630 &   0.086 \\
MAP     &0.227  &  0.194   &0.492 & 0.076\\
MRR    &  0.702  &  0.564 & 0.625& 0.115   \\
NDCG    & 0.636 &  0.560 & 0.656  & 0.137 \\
\bottomrule
\end{tabular}
\caption{Performance of U-NeuRec with pairwise training algorithm}
\vspace{-2mm}
\end{table}

\begin{figure}[h]
\begin{center}
\begin{minipage}[t]{4.0cm}
\includegraphics[width=4.0cm]{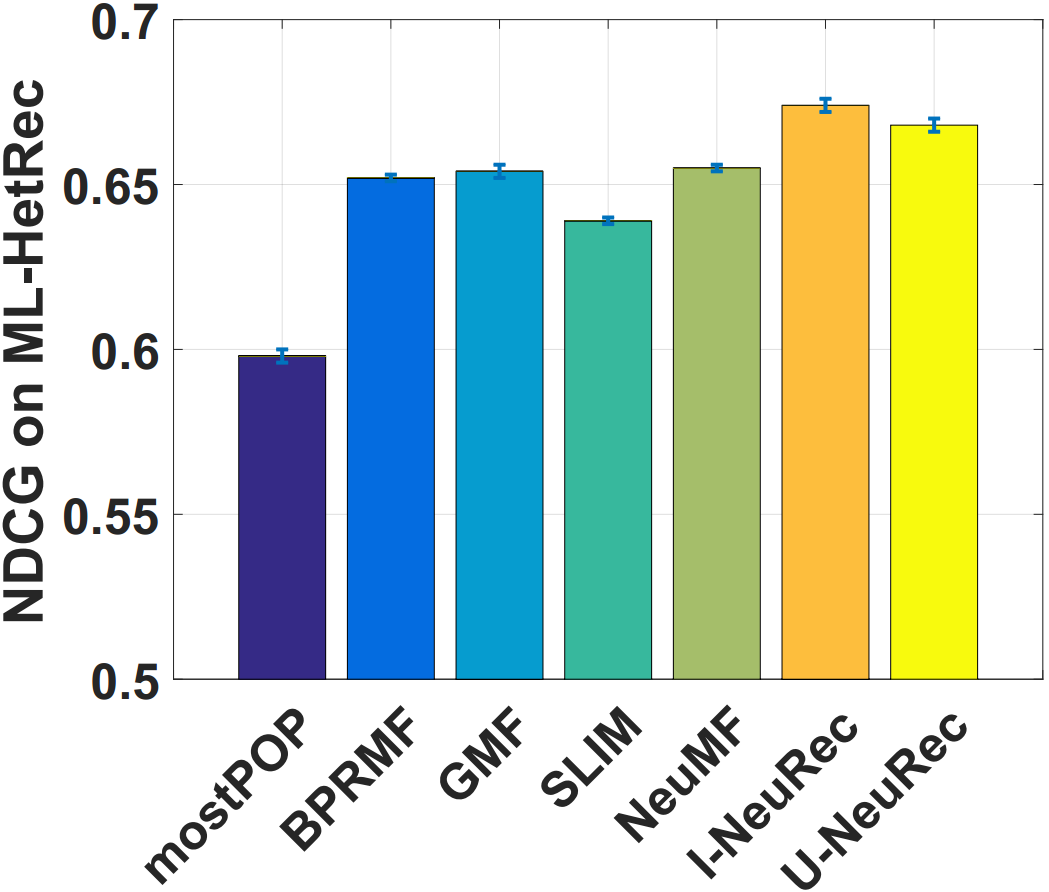}
\centering{(a)}
\end{minipage}
\begin{minipage}[t]{4.0cm}
\includegraphics[width=4.0cm]{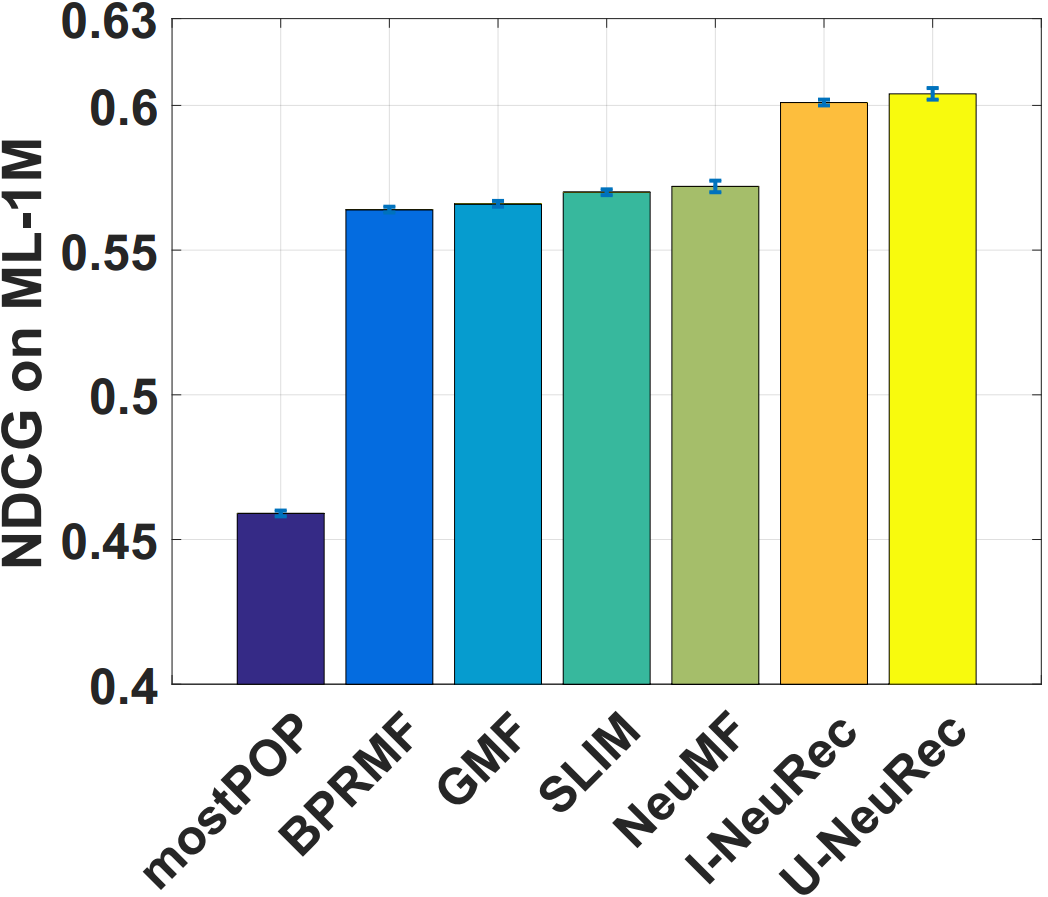}
\centering{(b)}
\end{minipage}
\begin{minipage}[t]{4.0cm}
\includegraphics[width=4.0cm]{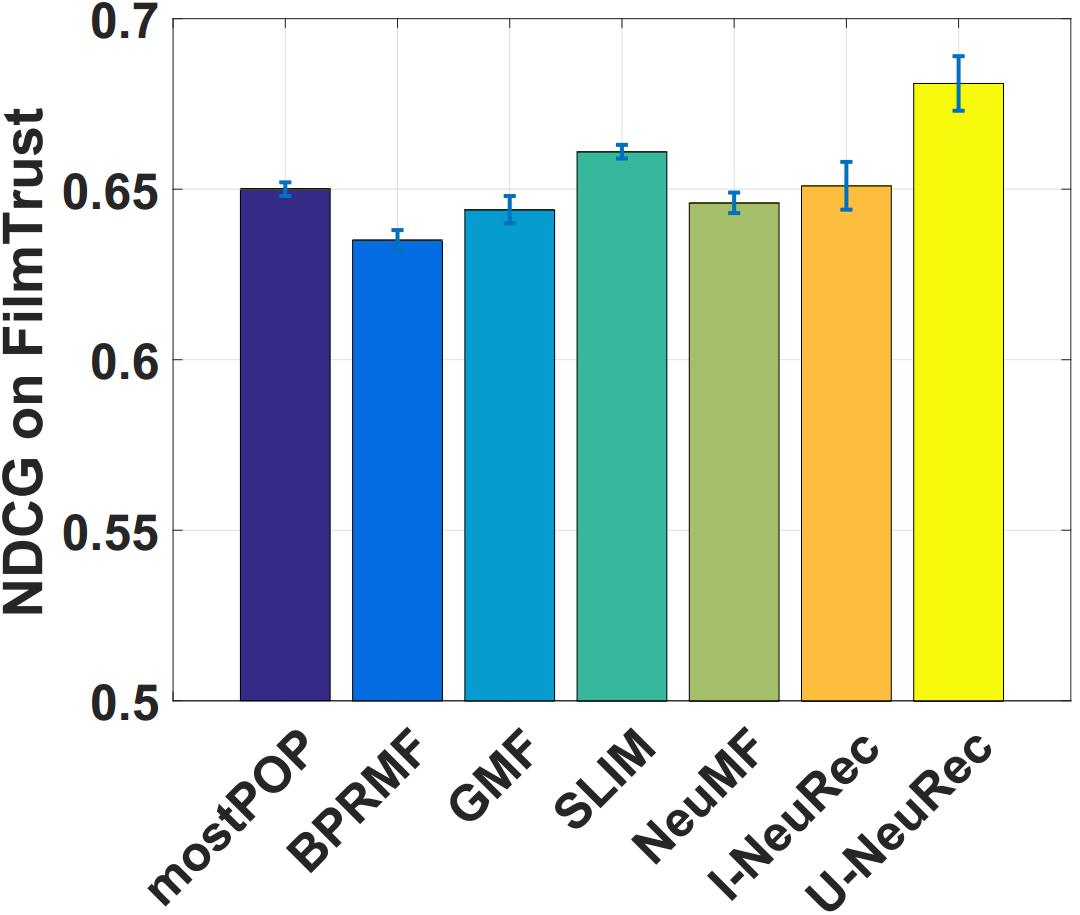}
\centering{(c)}
\end{minipage}
\begin{minipage}[t]{4.0cm}
\includegraphics[width=4.0cm]{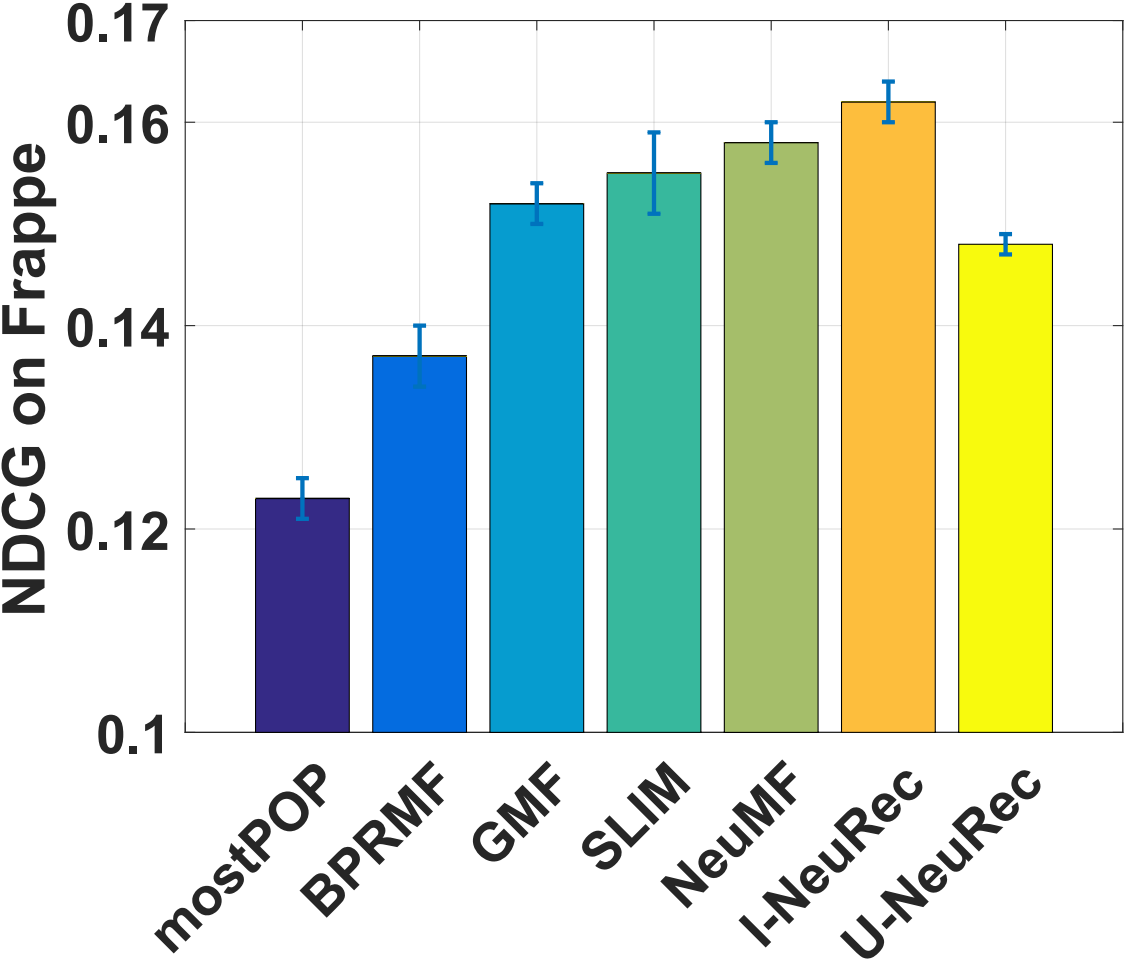}
\centering{(d)}
\end{minipage}
\caption{ NDCG Comparison on dataset (a) Movielens HetRec; (b) Movielens 1M; (c) FilmTrust; (d) Frappe.}
\label{fig:observation}
\end{center}
\vspace{-5mm}
\end{figure}


\begin{table}[h]
\centering

\label{my-label}
\begin{tabular}{c|cccc}
\toprule
 & \multicolumn{1}{c}{ML HetRec}     & \multicolumn{1}{c}{ML 1M}       & \multicolumn{1}{c}{FilmTrust}   & \multicolumn{1}{c}{FRAPPE}   \\
\midrule
P@5     & 0.415 &0.345  &  0.413& 0.039 \\
P@10    &  0.394  &    0.304   &  0.346    &  0.036\\
R@5     &   0.036 &  0.075  &  0.397   &  0.037  \\
R@10    & 0.066  & 0.127   &  0.618  &   0.063  \\
MAP     & 0.210 &  0.193   & 0.483  &  0.063\\
MRR    &  0.579   &   0.554    & 0.610 &  0.108  \\
NDCG    & 0.615 &    0.556  &  0.644  & 0.129 \\
\bottomrule
\end{tabular}
\caption{Performance of I-NeuRec with pairwise training algorithm}
\vspace{-2mm}
\end{table}

\subsection{Sensitivity to Neural Network Parameters }
In the following text, we systematically investigate the impacts of neural hyper-parameters on U-NeuRec with dataset FilmTrust (I-NeuRec has a similar pattern to U-NeuRec). In each comparison, we keep other settings unchanged and adjust the corresponding parameter values.
\begin{figure}[t]
\begin{center}
\begin{minipage}[t]{4.2cm}
\includegraphics[width=4.2cm]{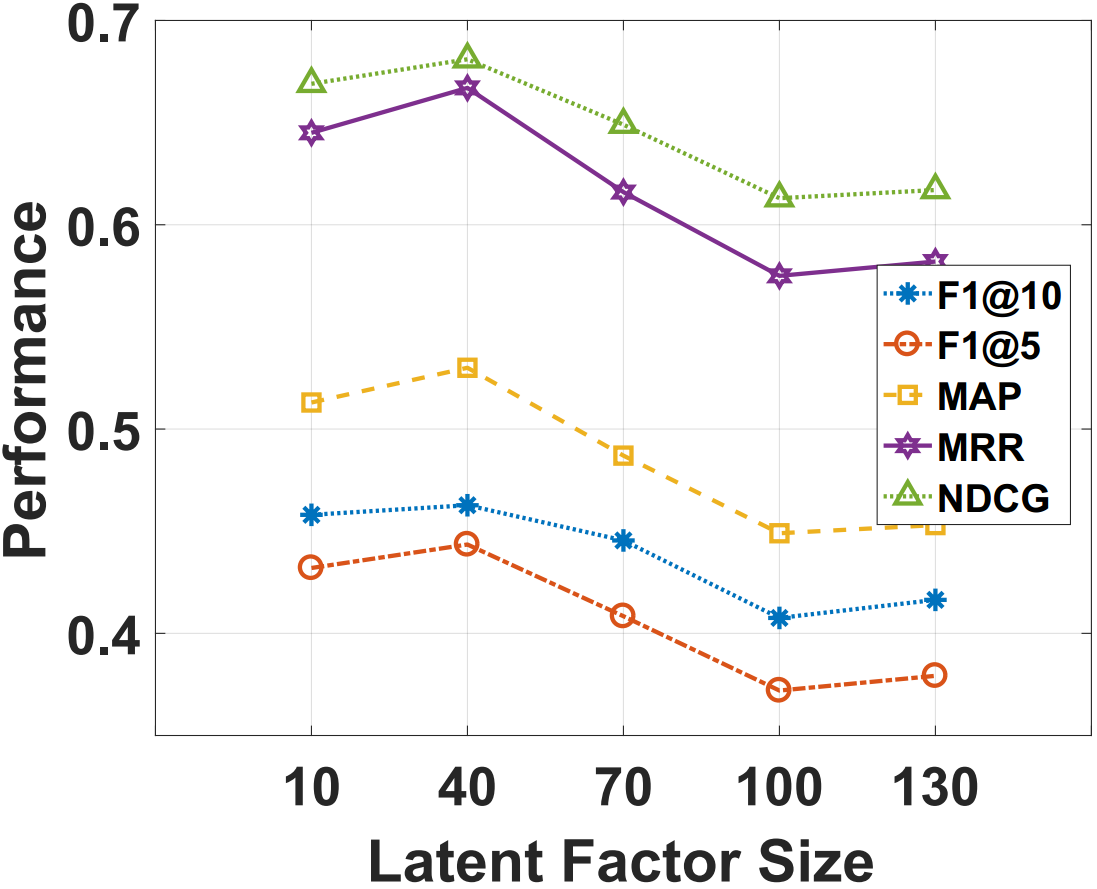}
\centering{(a)}
\end{minipage}
\begin{minipage}[t]{4.2cm}
\includegraphics[width=4.2cm]{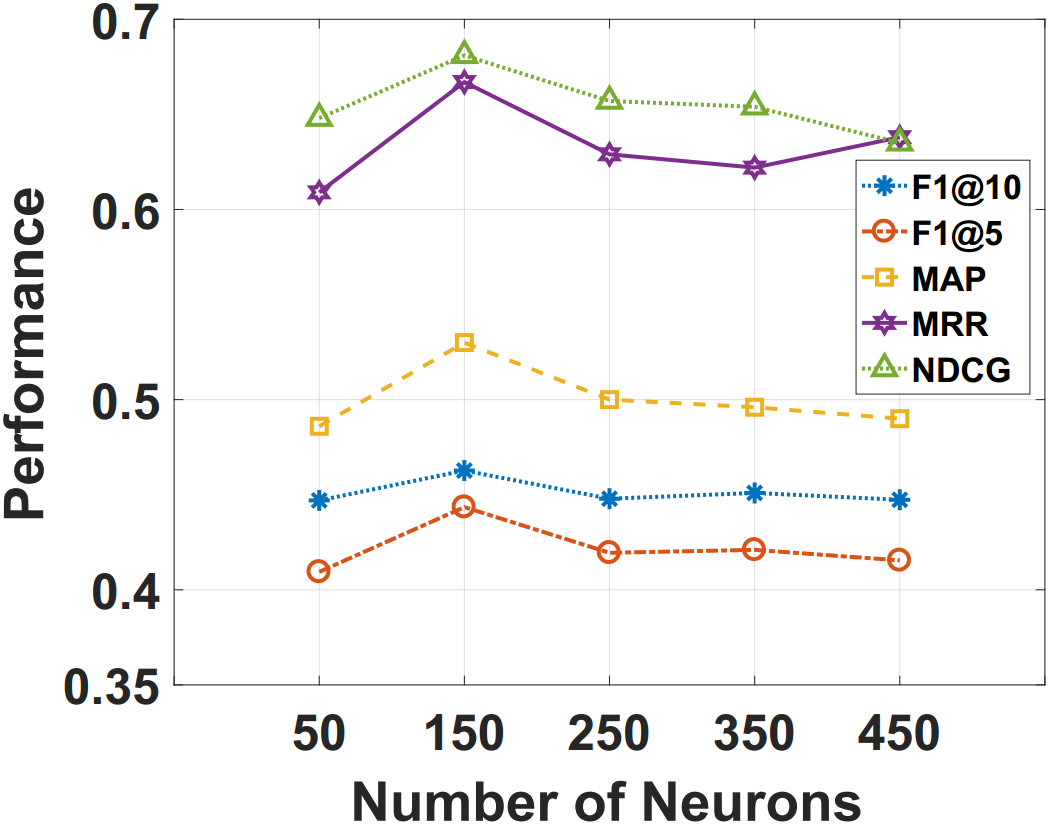}
\centering{(b)}
\end{minipage}
\begin{minipage}[t]{4.2cm}
\includegraphics[width=4.2cm]{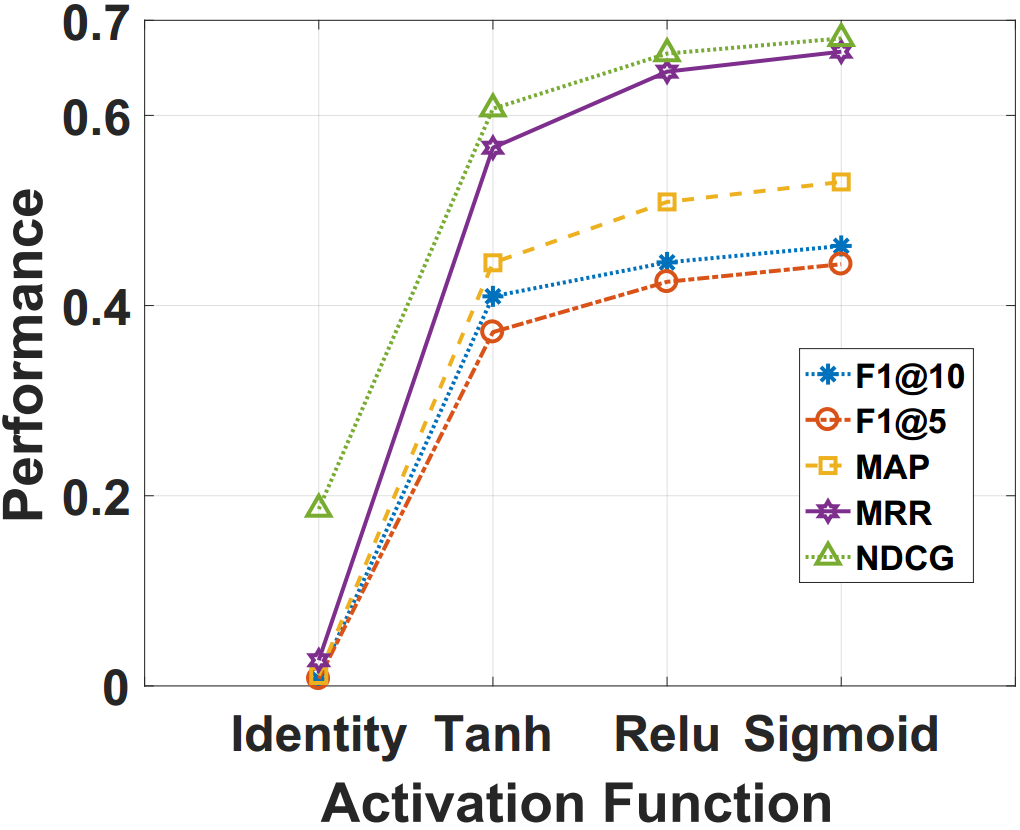}
\centering{(c)}
\end{minipage}
\begin{minipage}[t]{4.2cm}
\includegraphics[width=4.2cm]{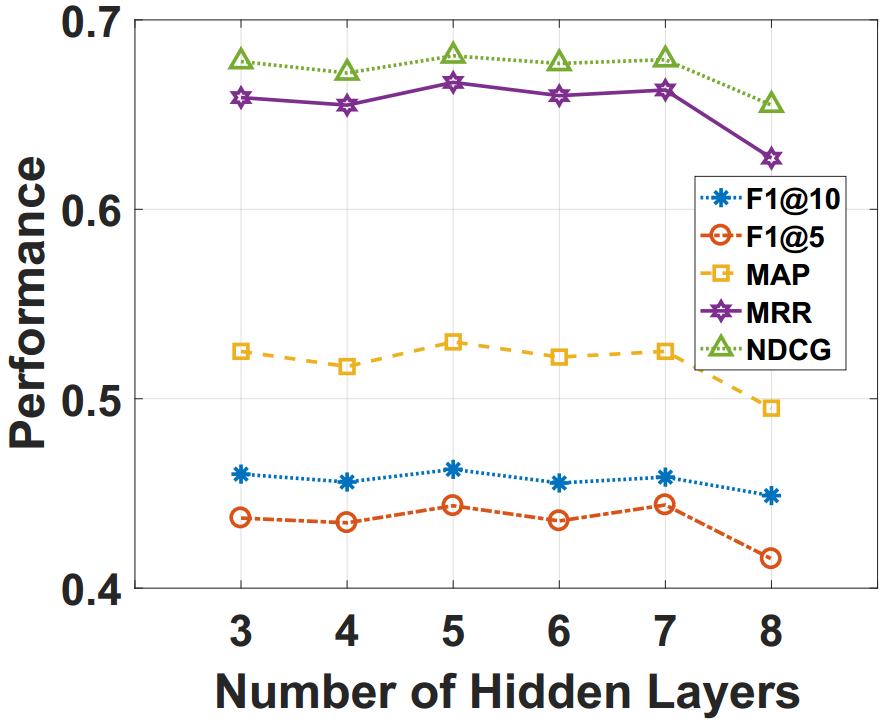}
\centering{(d)}
\end{minipage}
\caption{ Sensitivity of U-NeuRec to neural network hyper-parameter: (a) Latent Factor Size $k$; (b) Number of Neurons; (c) Activation Function; (d) Depth of Neural Network.}
\label{fig:observation}
\end{center}
\vspace{-5mm}
\end{figure}

\subsubsection{Latent Factor Size} Similar to latent factor model~\cite{koren2015advances}, the latent factor dimension poses great influence on the ranking performances. Larger latent factor size will not increase the performance and may even result in overfitting. In our case, setting $k$ to a value around $30$ to $50$ is a reasonable choice.

\subsubsection{Number of Neurons}
We set the neurons size to 50, 150, 250, 350 and 450 with a constant structure. As shown in Figure 3(b), both too simple and too complex model will decrease the model performance: simple model suffers from under-fitting while complex model does not generalize well on test data.


\subsubsection{Activation Function}
We mainly investigate activation functions: $sigmoid$, $tanh$, $relu$ and $identity$. We apply the activation function to all hidden layers. Empirically study shows that the $identity$ function performs poorly with NeuRec, which also demonstrates the effectiveness of introducing non-linearity. $sigmoid$ outperforms the other three activation functions. One possible reason is that $sigmoid$ can restrict the predicted value in range of $[0,1]$, so it is more suitable for binary implicit feedback.

\subsubsection{Depth of Neural Network}
Another key factor is the depth of the neural network. From Figure 3(d), we observe that our model achieves comparative performances with hidden layers number set to 3 to 7. However, when we continue to increase the depth, the performance drops significantly. Thus, we would like to avoid over-complex model by setting the depth to an appropriate small number.

\section{Conclusion and Future Work}
In this paper, we propose the NeuRec along with its two variants which provide a better understanding of the complex and non-linear relationship between items and users. Experiments show that NeuRec outperforms the competing methods by a large margin while reducing the size of parameters substantially. In the future, we would like to investigate methods to balance the performance of I-NeuRec and U-NeuRec, and incorporate items/users side information and context information to further enhance the recommendation quality. In addition, more advanced regularization techniques such as batch normalization could also be explored.


\bibliographystyle{named}
\bibliography{ijcai18}

\end{document}